\documentclass{aastex631}
\usepackage{float}

\shorttitle{Accretion Flows: Outflows VS Advection}
\shortauthors{Wu et al.}

\begin{document}

\title{Thermal Equilibrium Solutions of Black Hole Accretion Flows: 
Outflows VS Advection}

\author{Wen-Biao Wu}
\author{Wei-Min Gu}
\author{Mouyuan Sun}

\affiliation{Department of Astronomy, Xiamen University, Xiamen,
Fujian 361005, P. R. China; guwm@xmu.edu.cn}

\begin{abstract}
Observations and numerical simulations have shown that outflows generally
exist in the accretion process. We revisit the thermal equilibrium solutions
of black hole accretion flows by including the role of outflows.
Our study focuses on the comparison of the cooling rate of outflows with
that of advection.
Our results show that, except for the inner region, outflows can dominate
over advection in a wide range of the flow, which is in good agreement with
previous numerical simulations. We argue that an advection-dominated inner
region together with an outflow-dominated outer region should be a general
radial distribution for both super-Eddington accretion flows and
optically thin flows with low accretion rates.
\end{abstract}


\section{Introduction} \label{sec: intro}

Three well-known models have been widely investigated in black-hole accretion systems,
namely, the standard thin disk \citep[][hereafter SSD]{1973A&A....24..337S},
the slim disk \citep{1988ApJ...332..646A}, and the optically thin advection-
dominated accretion flow
\citep[][hereafter ADAF]{1994ApJ...428L..13N,1995ApJ...444..231N}.
The SSD is optically thick, geometrically thin, and radiative cooling dominated.
This model has been proposed to be central engines of  luminous active galactic
 nuclei (AGNs) and black hole X-ray binaries (BHXBs) in the high/soft state
 \citep[e.g.,][]{2008bhad.book.....K}. For the slim disk, it has a high mass 
accretion rate (super-Eddington), thus the radiation diffusion timescale is 
larger than the viscous timescale. This disk is optically thick and geometrically
slim, which may work for ultraluminous X-ray sources, bright microquasars, and 
narrow-line Seyfert 1 galaxies \citep[e.g.,][]{1999ApJ...516..420W,
2000PASJ...52..499M,2007A&ARv..15....1D,2017ARA&A..55..303K,
2018MNRAS.479.3978K}.
The ADAF is radiatively inefficient, where the viscous heating is
mainly balanced by the advective cooling.
The flow is geometrically thick and its gas 
temperature is close to the virial temperature.
This model has been applied to describe sources such as the supermassive black
hole in our Galactic center, Sagittarius
${\rm A}^{\ast}$ (Sgr~${\rm A}^{\ast}$),
low-luminosity AGNs, and BHXBs in the low/hard and quiescent states
\citep[e.g.,][]{2008NewAR..51..733N,2014ARA&A..52..529Y}. The above models
assume that the accretion rate remains constant during the accretion process.

Recent observations showed that outflows exist in the geometrically thin
disks \citep[e.g.,][]{2015ARA&A..53..115K,2016AN....337..368D,
2016ApJ...830L...5H}, the super-Eddington accretion disks \citep[e.g.,][]
{2009MNRAS.397.1836G,2011MNRAS.417..464M,2015ApJ...806...22D}, and the
radiatively inefficient accretion flows \citep[e.g.,][]{2013Sci...341..981W,
2016Natur.533..504C,2016ApJ...830L...5H,
2019MNRAS.483.5614M,2019ApJ...879L...4M}.
This means that the accretion rate is no longer a constant. Apart from
observations, outflows were also found in the hydrodynamical (HD) and
magnetohydrodynamical (MHD) numerical simulations of the geometrically thin
disks \citep[e.g.,][]{2011ApJ...736....2O,2016PASJ...68...16N,
2020MNRAS.494.3616N}, the super-Eddington accretion disks \citep[e.g.,][]
{2005ApJ...628..368O,2011ApJ...736....2O,2014ApJ...796..106J,
2014MNRAS.439..503S,2015MNRAS.453.3213S,2017PASJ...69...92K,
2018PASJ...70..108K,2020ApJ...888...86Z}, and the radiatively inefficient
accretion flows \citep[e.g.,][]{1999MNRAS.310.1002S,2012ApJ...761..129Y,2012MNRAS.426.3241N,
2012ApJ...761..130Y,2015ApJ...804..101Y}.

Outflows as a signature of the super-Eddington accretion were discussed by
\cite{1973A&A....24..337S}. In this pioneering paper, they supposed that
outflows are inevitable when the luminosity of the disk exceeds the Eddington
limit, since the radiation force is greater than the gravity.
More importantly, \cite{1978ApJ...221..652P} investigated the stability of
accretion disks and proposed that wind escaping from the disk surface can
have a stabilizing effect. Later, \cite{1981Natur.294..235A}
showed that the innermost parts of accretion disks are thermally
and secularly stable owing to the general relativistic effect, where
the physical picture is analogous to the case of Roche-lobe overflow
in close binaries. On the other hand,
\cite{1994ApJ...428L..13N,1995ApJ...444..231N} argued that
outflows are likely to occur because the Bernoulli parameter is positive
in some regions.
\cite{1999MNRAS.303L...1B} constructed the adiabatic inflow-outflow
solutions to describe outflows. They assumed that the accretion rate is a 
function of radius $\dot{M}\propto r^p$, where $p$ is in the range $[0,1]$.
\cite{2008ApJ...681..499X} also used a global method to show the influence
of the outflows on the disk structure by this relationship. Further numerical
simulations showed the power-law index $p$ in a range of [0.5, 1]
\citep[e.g.,][]{1999MNRAS.310.1002S,2005ApJ...628..368O,2012MNRAS.426.3241N,
2012ApJ...761..129Y,2018MNRAS.474.1206B}. In the case of Sgr $\rm{A}^{\ast}$,
however, a relatively low value for the index $p$ is preferred, such as
$p = 0.25$ \citep{1999ApJ...520..298Q}, 0.27
\citep{2003ApJ...598..301Y}, and 0.37 \citep{2019MNRAS.483.5614M}.
\cite{2011MNRAS.413.1623D} proposed a model for super-Eddington accretion
flows that the disk remains slim and a significant wind is accelerated.
Some works also demonstrated that the super-Eddington accretion
\citep{2007ApJ...660..541G,2015ApJ...799...71G,
2015MNRAS.448.3514C,2019ApJ...885...93F} and the optically thin ADAF
\citep{2015ApJ...799...71G} ought to have outflows.

In this work, we revisit the thermal equilibrium solutions of black hole
accretion flows by including the role of outflows. Our study focuses on the
comparison of the cooling rate of outflows with that of advection.
The paper is organized as follows. The basic equations for our model are
described in Section $\ref{sec: model}$. Numerical results and analyses are
shown in Section $\ref{sec: numerical Results}$. Conclusions and discussion
are presented in Section $\ref{sec:Summary}$.

\section{Basic Equations} \label{sec: model}

In this section, we describe the basic equations of our model. We consider a
steady state axisymmetric accretion flow, and use the pseudo-Newtonian
potential $\Phi = -GM_{\rm{BH}}/(R-R_{\rm{g}})$, where $M_{\rm{BH}}$ is the
mass of the black hole and $R_{\rm{g}}$ is the Schwarzschild radius. The
vertical scale height of the flow is $H = c_{\rm{s}}/\Omega_{\rm{K}}$, where
$\Omega_{\rm{K}}$ is Keplerian angular velocity, and $c_{\rm{s}} = (P/\rho)^{1/2}$
is the isothermal sound speed, with $P$ and $\rho$ being the pressure and mass
density, respectively. The kinematic viscosity coefficient is expressed as
$\nu = \alpha c_{\rm{s}} H$, where $\alpha$ is the constant viscosity parameter.

The basic equations describing the flow contain the continuity,
radial momentum, azimuthal momentum, and energy equations.
The continuity equation is
\begin{equation}
\frac{1}{R}\frac{d}{d R}\left(R\Sigma V_{\rm{R}}\right)+\frac{1}{2\pi R}\frac
{d \dot{M}_{\rm w}}{d R} = 0 \ , \label{eq1}
\end{equation}
where $\Sigma$ is the surface density defined as $\Sigma \equiv 2\rho H$,
and $V_{\rm{R}}$ is the radial velocity, which is defined to be negative when
the flow is inward. The outflow mass-loss rate $\dot{M}_{\rm{w}}$ is
 \citep{1999MNRAS.309..409K}
\begin{equation}
\dot{M}_{\rm w}(R) = \int_{R_{\rm{in}}}^{R} 4\pi R^\prime\dot{m}_{\rm w}
(R^\prime)dR^\prime \ , \label{eq2}
\end{equation}
where $R_{\rm{in}}$ denotes the radius at the inner edge of the disk and
$\dot{m}_{\rm w}$ is mass-loss rate per unit area from each disk face.

Due to the influence of outflows, we assume that the accretion rate
$\dot{M}$ varies with radius as follows \citep{1999MNRAS.303L...1B}:
\begin{equation}
\dot{M} =-2\pi R\Sigma V_{\rm{R}}=\dot{M}_{\rm{outer}}\left(\frac{R}
{R_{\rm{outer}}}\right)^{p} \ , \label{eq3}
\end{equation}
where $\dot{M}_{\rm{outer}}$ is the mass accretion rate at the outer boundary
$R_{\rm{outer}}$.

Some numerical simulations showed that outflows in super-Eddington accretion
cases and ADAFs are stronger than that in SSDs
\citep[e.g.,][]{2011ApJ...736....2O,2014SSRv..183..353O}, which can be
physically understood as follows.
The effective cooling of radiation in SSDs leads to a low
temperature of the disk, i.e., a negative Bernoulli parameter of the flow.
Thus, only relatively weak outflows may be produced by SSDs.
On the contrary, for super-Eddington accretion cases,
even though the temperature of the disk is only slightly higher than
that of SSDs, a large amount of photons trapped in the disk result in
high radiation pressure, which can contribute to strong outflows.
In addition, for ADAFs, the extremely high temperature of the disk due
to energy advection cause positive Bernoulli parameter
\citep[e.g.,][]{1994ApJ...428L..13N,1997ApJ...476...49N}, which can also contribute
to strong outflows. In summary, energy advection is helpful to produce
strong outflows, no matter the physics of advection is related to
photons or gas. It is known that the dimensionless thickness $H/R$
of the disk well describes the strength of advection, i.e., $H/R \ll 1$
for SSDs and $H/R \la 1$ for slim disks and ADAFs.
We therefore assume that the power-law index $p$ is proportional
to $H/R$ of the disk, i.e., $p = \lambda (H/R)$, where $\lambda$ is a
constant. In our opinion, such an assumption is more appropriate than
a fixed value of $p$ for different accretion models.

Using Equations~(\ref{eq1})-(\ref{eq3}), we obtain this relation,
\begin{equation}
\dot{m}_{\rm{w}}=\frac{\dot{M}p}{4\pi R^2} \ . \label{eq4}
\end{equation}
The integrated radial momentum equation and the azimuthal equation of motions
can be respectively written as 
\begin{equation}
V_{\rm R}\frac{dV_{\rm R}}{dR} + \left(\Omega_{\rm K}^2-\Omega^2\right)R + 
\frac{1}{\rho}\frac{dP}{dR} = 0 \ , \label{eq5}
\end{equation}
\begin{equation}
 -\frac{1}{R}\frac{d}{d R}\left(R^3\Sigma V_{\rm R}\Omega \right)+\frac{1}{R}
\frac{d}{d R}\left(R^3\nu \Sigma\frac{d \Omega}{d R}\right)-
\frac{\left(lR\right)^2\Omega}{2\pi R}\frac{d \dot{M_{\rm w}}}{d R} = 0 \ ,
\label{eq6}
\end{equation}
where the last term on the left-hand side of Equation~(\ref{eq6})
represents angular momentum carried by the outflowing materials.
Here, $l = 0$ corresponds to a non-rotating outflow, and $l = 1$
corresponds to the outflowing materials carrying away the specific
angular momentum at the point of ejection.
The cases with $l > 1$ correspond to centrifugally driven magnetic disk winds
that extract more angular momentum from the disk \citep{1999MNRAS.309..409K}.

The pressure $P$ is the sum of gas and radiation pressure:
\begin{equation}
P = \frac{\rho k_{\rm{B}}}{\mu m_{\rm{p}}} \left(T_{\rm i}+T_{\rm e}\right)+
\frac{Q_{\rm{rad}}}{4c}\left(\tau +\frac{2}{\sqrt{3}}\right) \ , \label{eq7}
\end{equation}
where $T_{\rm{i}}$ and $T_{\rm{e}}$ are the ion temperature and the electron
temperature, respectively, and $T_{\rm{e}} = \min \ (T_{\rm{i}},
6\times 10^9~{\rm K})$. $\mu = 0.617$ is the mean molecular weight,
and $\tau = (\kappa_{\rm{es}} + \kappa_{\rm{abs}})\rho H$
is the total optical depth, where $\kappa_{\rm{es}} =
0.34 \rm{\ cm^2 \ g^{-1}}$ and $\kappa_{\rm{abs}} = 0.27\times 10^{25} \rho 
T_{\rm{e}}^{-3.5}\ \rm{cm^2 \ g^{-1}}$ \citep[e.g.,][]{1996ApJ...471..762A}.

The energy equation is written as
\begin{equation}
Q_{\rm{vis}} = Q_{\rm adv} + Q_{\rm rad} + Q_{\rm w} \ , \label{eq8}
\end{equation}
where $ Q_{\rm{vis}}$, $Q_{\rm adv}$, and $ Q_{\rm rad}$ are the viscous
heating rate, the advective cooling rate, and the radiative cooling rate,
respectively. Their expressions are as follows,
\begin{equation}
Q_{\rm{vis}} = \nu \Sigma\left(R\frac{d\Omega}{dR}\right)^2 \ , \label{eq9}
\end{equation}
\begin{equation}
Q_{\rm adv} = \Sigma V_{\rm R} T\frac{ds}{dR} = \Sigma V_{\rm R}\left(\frac{1}
{\gamma-1}\frac{dc_{\rm s}^2}{dR} - \frac{c_{\rm s}^2}{\rho}\frac{d\rho}{dR}
\right) \ , \label{eq10}
\end{equation}
\begin{equation}
Q_{\rm rad} = 8\sigma T_{\rm e}^4\left(\frac{3\tau}{2} + \sqrt{3} + 
\frac{8\sigma T_{\rm{e}}^4}{Q_{\rm {br}}^-}\right)^{-1} \ . \label{eq11}
\end{equation}
Equation~(\ref{eq11}) is valid in both optically thin and optically
thick regimes \citep{1995ApJ...452..710N}. The bremsstrahlung cooling is
given by \citep[e.g.,][]{1995ApJ...438L..37A}
\begin{equation}
Q_{\rm {br}}^- = 1.24\times 10^{21} H\rho^2 T_{\rm e}^{1/2}~{\rm erg~s^{-1}
cm^{-2}} \ . \label{eq12}
\end{equation}
The quantity $Q_{\rm w}$ in Equation~(\ref{eq8}) represents the energy
taken away by outflows, which is expressed as
\begin{equation}
Q_{\rm w} = 2f\eta \dot{m}_{\rm w} V_{\rm K}^2 \ , \label{eq13}
\end{equation}
where a factor of 2 represent the outflow energy is emitted from both sides of
the accretion disk.
$\eta$ is an outflow energy parameter and $V_{\rm K}$ is the Keplerian
velocity. By using Equation~(\ref{eq1}) and integrating
Equation~(\ref{eq6}), we have
\begin{equation}
\nu \Sigma = \frac{\dot{M}fg^{-1}}{3\pi}\left(1-\frac{l^2p}{p+\frac{1}{2}}
\right) \ , \label{eq14}
\end{equation}
where $g = -\left(2/3\right)\left(d \ln \Omega_{\rm{K}}/d \ln R\right)$ and
the factor $f = 1-[\Omega \left(3R_{\rm{g}}\right)/\Omega (R)]\left(3R_{\rm{g}}
/R\right)^{p+2}$. For $p = 0$, Equation~(\ref{eq14})
returns to Equation~(2.1) of \citet{1995ApJ...443L..61C}.

Following some previous works \citep[e.g.,][]{1994ApJ...428L..13N,
2000ApJ...540L..33G}, we adopt the self-similar assumptions and set
$\gamma = 1.5$.
Then Equations~(\ref{eq9})-(\ref{eq11}) are reduced to the following
algebraic forms:
\begin{equation}
\frac{1}{2}V_{\rm{R}}^2+\frac{5}{2}c_{\rm s}^2 + \left(\Omega^2-
\Omega_{\rm K}^2\right)R^2 = 0 \ , \label{eq15}
\end{equation}
\begin{equation}
Q_{\rm{vis}} = \frac{3\dot{M}\Omega^2fg}{4\pi}\left(1-\frac{l^2p}{p+\frac{1}
{2}}\right) \ , \label{eq16}
\end{equation}
\begin{equation}
Q_{\rm adv} = \frac{1}{4\pi}\frac{\dot{M}c_{\rm s}^2}{R^2} \ . \label{eq17}
\end{equation}
Finally, substituting Equation~(\ref{eq4}) into Equation~(\ref{eq13}), 
we have
\begin{equation}
Q_{\rm w} =\frac{f\eta p\dot{M}\Omega_{\rm{K}}^2}{2\pi} \ . \label{eq18}
\end{equation}
By solving the five equations, Equations~(\ref{eq3}),
(\ref{eq7}-\ref{eq8}), and (\ref{eq14}-\ref{eq15}), for the five variables
$\rho$, $T$, $c_{\rm s}$, $\Omega$, and $V_{\rm R}$ with given parameters
$M_{\rm BH}$, $\alpha$, $\dot M$, and $l$, we obtain the thermal equilibrium
solutions of accretion flows. In the following calculations, we fix
$M_{\rm{BH}} = 10M_{\sun}$, $\alpha = 0.1$, and $l = 1$.

\section{Numerical Results} \label{sec: numerical Results}

\begin{figure}
\centering
\includegraphics[width=12cm]{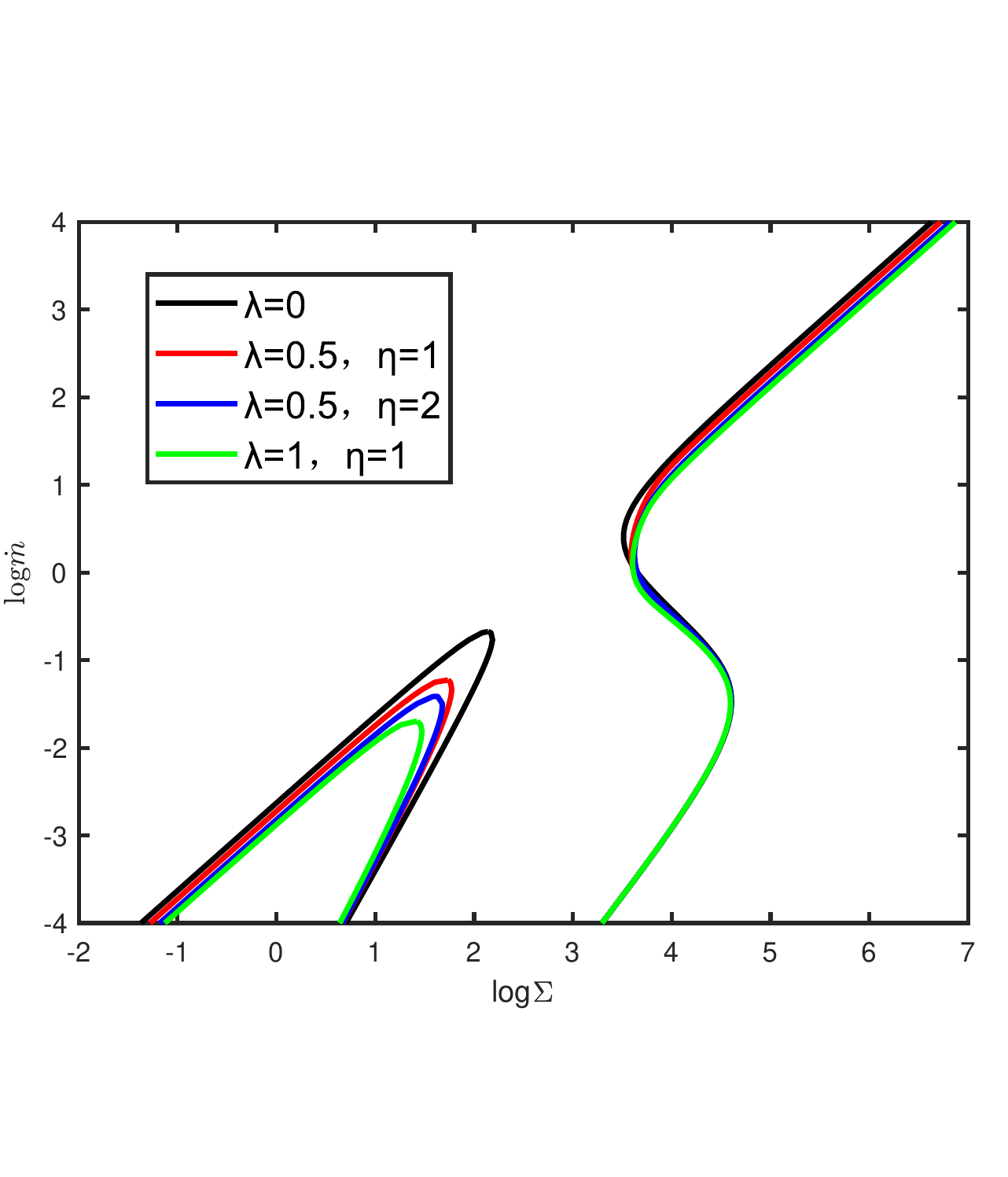}
\caption{Thermal equilibria of accretion disks at $R = 10 R_{\rm g}$ for
various $\lambda$ and $\eta$. The color lines correspond to the cases for
($\lambda=0$, black), ($\lambda=0.5, \eta=1$, red), ($\lambda=0.5, \eta=2$,
blue), and ($\lambda=1, \eta=1$, green). \label{fig:fig1}}
\end{figure}

In this section, we describe the numerical results of the accretion flows with
outflows. Figure~\ref{fig:fig1} shows thermal equilibrium solutions at 
$R = 10\ R_{\rm{g}}$ in the $\log \dot{m}-\log \Sigma$ plane, where
 $\dot{m}$ is the accretion rate normalized by the Eddington accretion rate
$\dot{M}_{\rm{Edd}} = 64\pi GM_{\rm{BH}}/c\kappa_{\rm{es}}$. The black line
represents the solutions under the no-outflow assumptions \citep[e.g.,][]
{1995ApJ...438L..37A,1995ApJ...443L..61C,1998ApJ...505L..19T,
2000ApJ...540L..33G}. The curve on the left is composed of two branches, of
which the upper one is for ADAFs and the lower one is for SLE disks
\citep{1976ApJ...204..187S}. The right S-shaped curve is composed of three
branches, of which the upper one is for slim disk, the middle one for radiation
pressure-supported SSDs, and the lower one for gas pressure-supported SSDs.
The red, blue and green lines represent the results with
($\lambda=0.5, \eta=1$), ($\lambda=0.5, \eta=2$), and ($\lambda=1, \eta=1$),
respectively.
It is seen from the figure that the maximal accretion rate of the left curve
decreases with increasing cooling effects of outflows.
However, the change of the right S-shaped curve is quite slight.
The obtained values of $p$ of ADAFs and slim branches are
$p \sim 0.23$ for $\lambda = 0.5$ and $p \sim 0.39$ for $\lambda = 1$.
For comparison, we have $p \sim 0.003$ for gas-pressure-supported SSDs.
For the ADAF branch, the values of $p$ are in good consistent with
the fitting results of the Sgr~$\rm{A}^{\ast}$ observations under
radiatively inefficient accretion model
\citep[e.g.,][]{1999ApJ...520..298Q,2003ApJ...598..301Y,
2019MNRAS.483.5614M}.

\begin{figure}
\centering
\includegraphics[width=12cm]{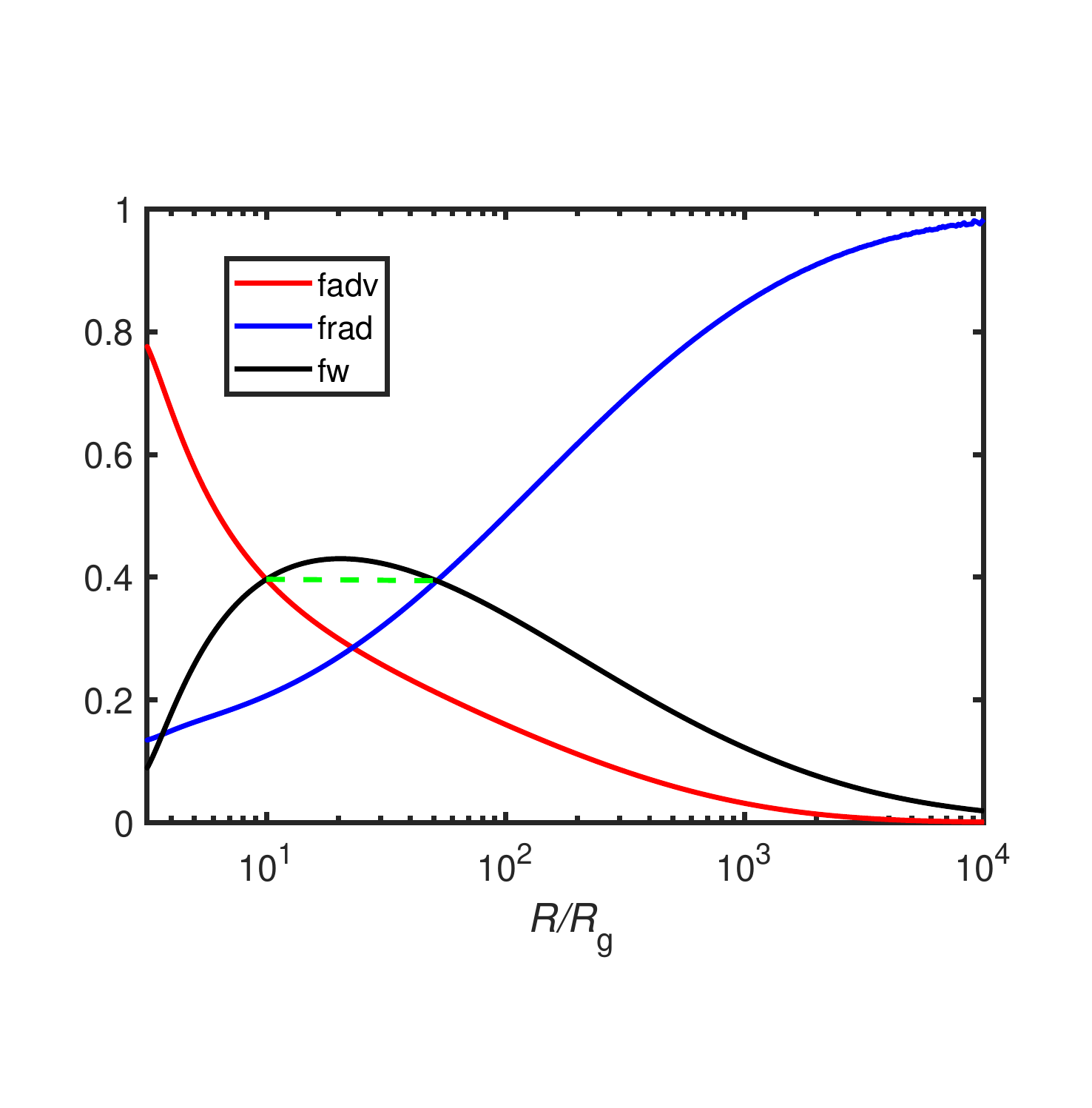}
\caption{Variations of $f_{\rm{adv}}$ (red line), $f_{\rm{rad}}$ (blue line),
$f_{\rm{w}}$ (black line) with $R$ for $\dot{m} = 100, \eta = 1,
\lambda = 0.5$.
The green dashed line denotes the outflow-cooling-dominated region 
($f_{\rm{w}} > f_{\rm{adv}}$, $f_{\rm{w}} > f_{\rm{rad}}$).
\label{fig:fig2}}
\end{figure}

\begin{figure}
\centering
\includegraphics[width=18cm]{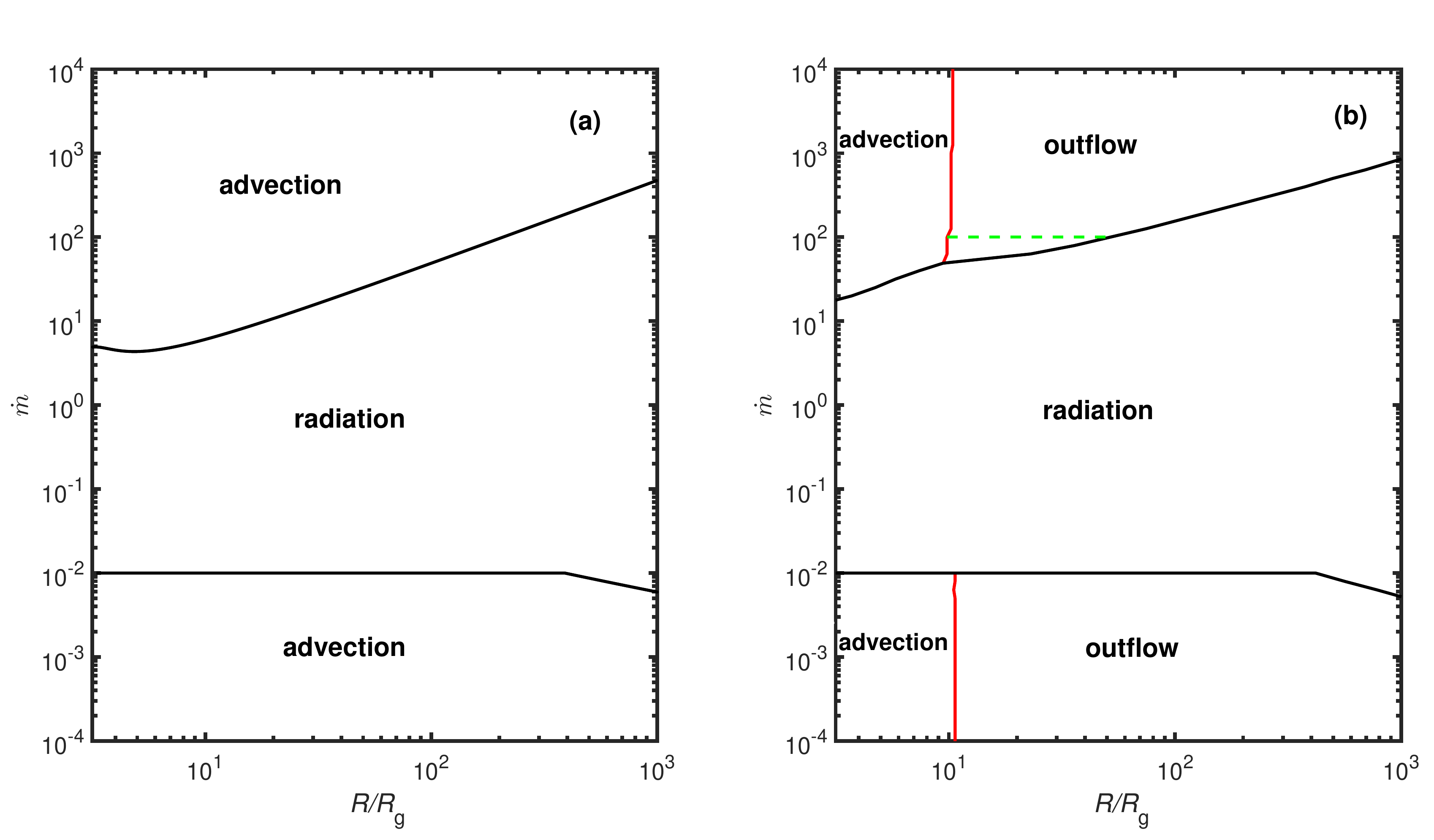}
\caption{Distribution of thermal equilibrium solutions for $\eta = 1$ and
$\lambda = 0.5$.
In panel (a), the parameter space is divided into three regions by two curves.
The upper region, denoted as ``advection" ($f_{\rm{adv}} > f_{\rm{rad}}$), 
corresponds to the super-Eddington accretion flows, and the lower ``advection"
region represents the ADAFs. The solutions of panel (a) are under the 
no-outflow assumptions. On the contrary, panel (b) takes outflows into account.
The ``advection" region corresponds to ($f_{\rm{adv}} > f_{\rm{rad}}$,
$f_{\rm{adv}} > f_{\rm{w}}$), the ``outflow" region corresponds to 
($f_{\rm{w}} > f_{\rm{adv}}$, $f_{\rm{w}} > f_{\rm{rad}}$), and 
the ``radiation" region corresponds to ($f_{\rm{rad}} > f_{\rm{adv}}$, 
$f_{\rm{rad}} > f_{\rm{w}}$). The two red lines ($R_{\rm b}$) represent the 
boundary between outflow-dominated and advection-dominated cases.
The green dashed line corresponds to the example solution 
for $\dot{m} = 100$ in Figure~\ref{fig:fig2}.
\label{fig:fig3}}
\end{figure}

In order to quantitatively understand the influence of outflows at different
radius, we fix $\dot{m} = 100$, $\eta = 1$, $\lambda = 0.5$, and
$R_{\rm outer} = 10^4 R_{\rm g}$. We then obtain variations of
$f_{\rm{adv}}(\equiv Q_{\rm{adv}} / Q_{\rm{vis}})$, $f_{\rm{rad}}
(\equiv Q_{\rm{rad}}/Q_{\rm{vis}})$ and $f_{\rm{w}}
(\equiv Q_{\rm{w}}/Q_{\rm{vis}})$ with radius. It is seen from
Figure~\ref{fig:fig2} that, in the inner regions cooling is dominated by
advection ($f_{\rm{adv}}$, the red line),
and in the middle regions cooling is dominated by outflows
($f_{\rm{w}}$, the black line).
Since the gravitational force in
the inner regions is greater than the radiation force,
and the radial velocity is large (i.e., the viscous timescale is short), only
a weak outflow can form. In the middle regions (denoted by the green dashed 
line), however, the situation is reversed as outflow dominance.
Such an outflow-dominated region varies significantly with varying
accretion rates, which is presented in Figure~\ref{fig:fig3}.

A main purpose of this work is to compare the cooling effects of outflows
with that of advection.
Figure~\ref{fig:fig3} is a description of thermal equilibrium
solutions of accretion flows with outflows in the $\dot{m}-R$ plane.
For a comparison, Figure~\ref{fig:fig3}(a) is under the no-outflow assumptions.
The $\dot{m}-R$ plane is divided into three regions by two curves.
The region above the upper curve ($f_{\rm adv} = 1/2$) represents
advective-cooling-dominated solutions ($f_{\rm adv} > f_{\rm rad}$) owing to
the photon trapping, which corresponds to the super-Eddington accretion cases.
The middle region represents solutions with cooling dominated by 
radiation ($f_{\rm rad} > f_{\rm adv}$).
The region below the lower curve also represents
advective-cooling-dominated solutions ($f_{\rm adv} > f_{\rm rad}$),
but the physics of advection is related to the internal energy
of accreted gas, which corresponds to the ADAF cases.
Here, the maximal critical mass accretion rate $\dot{m}_{\rm crit}$ of ADAF
is assumed to be around $0.01\dot{M}_{\rm Edd}$
($\dot{m}_{\rm crit} \sim \alpha^2 \dot{M}_{\rm{Edd}}$) for the inner part
with $R \la R_{\rm tr}$, where $R_{\rm tr}$ is around $10^2-10^3 R_{\rm g}$
for $\alpha = 0.1$. For $R > R_{\rm tr}$, $\dot{m}_{\rm crit}$ 
decreases with increasing $R$
\citep[see, e.g., Figure~8 of][]{1998tbha.conf..148N}.

Similar to Figure~\ref{fig:fig2}, Figure~\ref{fig:fig3}(b) is also
for $\eta = 1$ and $\lambda = 0.5$, which shows that there exists
a boundary radius $R_{\rm b}$ (red lines) which separates
an inner ``advection" region and an outer ``outflow" region.
By comparing Figures~\ref{fig:fig3}(a) and \ref{fig:fig3}(b), it is seen
that for $R > R_{\rm b}$, the original region with cooling dominated by
advection is replaced by a region with cooling dominated by outflows.
The green dashed line denotes the outflow-dominated region of the
example solution with $\dot{m} = 100$ in Figure~\ref{fig:fig2}.
It is seen from Figure~\ref{fig:fig3}(b) that
the outflow-dominated region increases with increasing accretion rates
for the super-Eddington accretion cases.
In addition, for the optically thin cases, Figure~\ref{fig:fig3}(b) shows
that a relatively wide outflow-dominated region exists for ADAFs,
and this region increases with decreasing accretion rates.
Thus, we argue that an advection-dominated inner region together with
an outflow-dominated outer region should be
a general radial distribution for both super-Eddington accretion flows
and optically thin flows with low accretion rates.

\begin{figure}
\centering
\includegraphics[width=12cm]{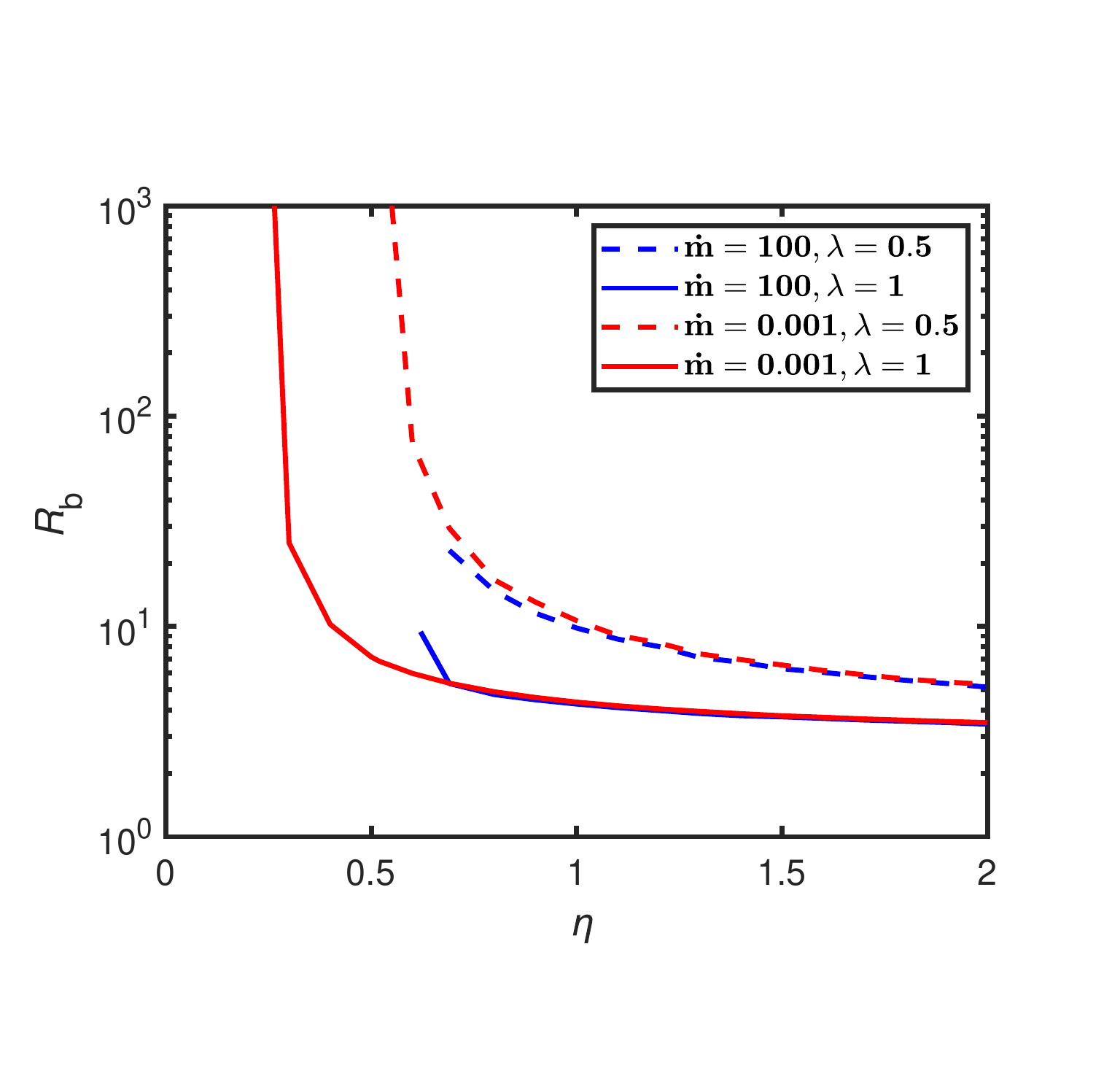}
\caption{Variations of the boundary radius $R_{\rm b}$ with $\eta$ for a
typical ADAF with $\dot m = 0.001$ and a typical slim disk with $\dot m = 100$.
The solid and dashed curves correspond to the cases with $\lambda = 1$
and $\lambda = 0.5$, respectively. \label{fig:fig4}}
\end{figure}

Figure~\ref{fig:fig4} shows the variation of $R_{\rm{b}}$
with the parameter $\eta$. The red solid (dashed) line represents
solutions with $\dot{m}=0.001$ and $\lambda = 1$ ($\lambda = 0.5$),
where $p$ is less than 0.4 (0.25).
The blue solid (dashed) line represents solutions with
$\dot{m}=100$ and $\lambda = 1$ ($\lambda = 0.5$), where $p$ is less than
0.33 (0.22). Since the temperature of ADAFs is close to the virial value,
the outflows are stronger and therefore the boundary $R_{\rm b}$ 
has a larger span in the radial direction.
In addition, the figure shows that, for a certain $\lambda$,
when $\eta \ga 1$, the boundary $R_{\rm b}$ is located between
the inner stable circular orbit ($3R_{\rm g}$) and $\sim 10R_{\rm g}$.
The physical reason is that the gravitational force in the inner region
of is very strong and therefore the outflows are restrained.
In other words, outflows from inside $\sim 10R_{\rm g}$ are quite weak,
which is consistent with previous numerical simulations
\citep[e.g.,][]{2012ApJ...761..130Y}.

\section{Conclusions and Discussion} \label{sec:Summary}

In this work, we have revisited the thermal equilibrium solutions of 
black-hole accretion flows by including the role of outflows. By comparing
the cooling rate of outflows with that of advection,
we have found that advection is important only in the inner regions
and outflows play a key role in balancing the viscous heating in the 
outer regions (Figure~\ref{fig:fig3}). We argue that an advection-dominated
inner region together with an outflow-dominated outer region should be
a general radial distribution for both super-Eddington accretion flows
and optically thin flows with low accretion rates.
In addition, we have also obtained the boundary
$R_{\rm b}$ as a function of $\dot{m}$ and $\eta$.
Our results are well physically understood, and 
well agree with observations and numerical simulations.

The present work is based on $\alpha = 0.1$ and $l = 1$. However,
there exists a critical viscosity parameter $\alpha_{\rm{crit}}$
for the structure of thermal equilibrium solutions
\citep{1995ApJ...443L..61C}.
For $\alpha > \alpha_{\rm{crit}}$, a new topological type of equilibria
appears where the ADAF branch can smoothly connect the slim disk branch.
Similarly, critical viscosity parameters should also exist if the effects
of outflows are taken into account. Thus, a new topological type of equilibria
may also exist for large values of $\alpha$.
In addition, the present study is based on $l = 1$. In fact,
outflows can extract more angular momentum from the disk, which
corresponds to $l > 1$, such as the centrifugally driven magnetohydrodynamic
winds \citep{1982MNRAS.199..883B}. 
This means more energy carried away by the outflows,
which may have significant effects on the disk structure.

A fundamental difference between neutron stars and black holes is that
the former has a hard surface whereas the latter has an event horizon.
For the ADAF case, most of the energy due to viscous heating process
in the flow is finally absorbed by the central black hole. However,
if the central object is a neutron star, the main energy generated by the
viscous process will be finally released around the hard surface.
Following this spirit, the observational evidence of event horizon was
previously proposed by comparing the radiation of quiescent state of
BHXBs and neutron star X-ray binaries
\citep[see, e.g., Figure 1 of][]{1997ApJ...478L..79N}.
\citet{1998AIPC..431..351L} proposed to compare quiescent neutron star
and black hole transients at similar mass accretion rates, and the best way
is to plot luminosities as a function of the binary orbital period.
By comparable orbital periods, the luminosities of
black hole systems are 2-3 orders of magnitude fainter than neutron star
systems \citep[see, e.g., Figures 4 and 5 of][]{2004ApJ...615..402M}.
We would point out that outflows may have effects on this issue,
which was not taken into account.
It is known that the inner region of a disk has essential contribution
to the total radiation.
According to our results, since the outflows in the inner region is quite
weak, in our opinion, the effects of outflows on this issue may not
be significant.

\begin{acknowledgments}
This work was supported by
the National Key R\&D Program of China under grant 2021YFA1600401,
and the National Natural Science Foundation of China under grants 11925301,
12033006, and 11973002.
\end{acknowledgments}

\end{document}